\documentclass[twocolumn]{article}
\usepackage[modulo,switch]{lineno}
\modulolinenumbers[1]

\usepackage[utf8]{inputenc}
\usepackage{amsmath}
\usepackage{amssymb}
\usepackage{graphicx}

\makeatletter\@ifundefined{date}{}{\date{}}
\makeatother

\evensidemargin\oddsidemargin \hoffset-22.4mm \voffset-28.4mm
\begin{document}

\title{Mathematical Analysis of the Probability of Spontaneous Mutations in HIV-1 Genome and Their Role in the Emergence of Resistance to Anti-Retroviral Therapy}

\author{Eslam Abbas$^{1}$ \\
$^{1}$ MBChB, Kobri El Koba Medical Complex, El Khalifa El Maamoun St.\\
   Intersection of El Fangary St., Heliopolis, Cairo, 11766, Egypt.\\}

\maketitle\thispagestyle{empty}

\begin{abstract}
\textbf{Background:} High mutability of HIV is the driving force of antiretroviral drug resistance, which represents a medical care challenge.\\
\\
\textbf{Method and Model Equation:} To detect the mutability of each gene in the HIV-1 genome; a mathematical analysis of HIV-1 genome is performed, depending on a linear relation wherein the probability of spontaneous mutations emergence is directly proportional to the ratio of the gene length to the whole genome length.
\begin{equation*}
{P_g}{S_i} =\frac{g}{G}
\end{equation*}\\
\\
\textbf{Results:} \textbf{tat}, \textbf{vpr} and \textbf{vpu} are the least mutant genes in HIV-1 genome. Protease \textbf{PROT} gene is the least mutant gene component of polymerases \textbf{pol}.\\
\\
\textbf{Conclusion:} \textbf{tat}, \textbf{vpr} and \textbf{vpu} are the best candidates for HIV-1 recombinant subunit vaccines or as a part of \textit{“prime and boost”} vaccine combinations. Also; the protease inhibitor-based regime represents a high genetic barrier for HIV to overcome.
\end{abstract}

\section*{INTRODUCTION:}

Retroviruses depend on their genetic instability as an evolutionary advantage to boost adaptive mutations. HIV has a very high genetic variability, which is a result of its fast replication cycle coupled with a high mutation rate [1]. HIV is capable of rapidly responding to the selective pressures imposed by the immune system and antiretroviral drugs. Drugs target only specific molecules, which are almost always proteins. Because the drug is so specific, any mutation in these molecules will interfere with or negate its effect, resulting in drug resistance [2].\\

Biological systems are paramount examples of complex dynamical systems, so mutation emergence is a fundamental property. The ability to calculate the probability of spontaneous mutations in a specific gene, will help provide an overview of the possibility of emergence of resistance to the protein translated from that gene during antiviral drug development. Also, this ability will be beneficial during protocoling of combination therapy.

\section*{MODEL EQUATION:}

The method used to formulate the mutability map of HIV-1 genetic pool, is a linear relation in which the probability of spontaneous mutation emergence, is directly proportional to the ratio of the gene length to the whole genome length.
\begin{equation*}
{P_g\propto{\frac{g}{G}}}
\end{equation*}
And so;
\begin{equation*}
{P_g}{S_i} = \frac{g}{G}
\end{equation*}
Wherein;\\
\({P_g}\) is the probability of spontaneous mutation emergence in a gene per duplicate.\\
\(\frac{g}{G}\) is the ratio between the gene length and the whole genome length.\\
\({S_i}\) is the stability index, which is a genome specific fixed value, and so:
\begin{equation*}
{P_G}{S_i}=1
\end{equation*}
\({P_G}\) is the probability of spontaneous mutation emergence in the genome per duplicate.\\
And so; the stability index represents the degree of stability of a genome.\\

The mutation rate of HIV-1 is approximately \(3\times10^{-5}\) per nucleotide base per cycle of replication [3]. The HIV-1 genome contains 9181 bases [4] and accordingly; the stability index of the HIV-1 genome is:
\begin{equation*}
{S_i}=\frac{1}{P_G}
\end{equation*}
So;
\begin{equation*}
{S_i}\approx3.63
\end{equation*}
\section*{RESULTS:}
A mathematical analysis, using the proposed equation, is performed and data are collected in 3 tables. Table 1 describes the analysis of the HIV-1 genetic pool, which indicates that the probability of spontaneous mutation emergence is lesser for \textbf{tat}, \textbf{vpr} and \textbf{vpu}. Table 2 describes the detailed analysis of the probability of spontaneous mutations emergence in the components of \textbf{pol} gene: reverse transcriptase, integrase and protease genes. These genes are translated into the main target proteins of antiretroviral therapy. The analysis indicates that reverse transcriptase \textbf{RT} is the most mutant gene of the polymerases and protease \textbf{PROT} is the least. Table 3 describes the analysis of the structural genes of \textbf{gag} and \textbf{env}, which indicates that \textbf{gp120} is more susceptible to spontaneous mutations emergence, and so has a higher diversity, than \textbf{gp41}.\\
\\
\subsubsection*{Table 1:} Analysis of the probability of spontaneous mutation in HIV-1 genetic pool; wherein \(g\) is the gene length, \(P_g\) is the probability of spontaneous mutation emergence in a specific gene and \(\%P_G\) is \% of spontaneous mutation probability of the genome.\\
\\
\begin{center}\begin{tabular}{|c|c|c|c|}
\hline
Gene & \((g)\) & \({P_g}\) & \% \({P_G}\)\\
\hline
gag & \(1502\) & \(\approx0.049\) & \(16.36 \%\)\\
\hline
pol & \(3011\) & \(\approx0.098\) & \(32.8 \%\)\\
\hline
vif & \(569\) & \(\approx0.017\) & \(6.2 \%\)\\
\hline
vpr & \(236\) &  \(\approx0.008\) & \(2.57 \%\)\\
\hline
tat & \(259\) & \(\approx0.008\) & \(2.9 \%\)\\
\hline
rev & \(349\) & \(\approx0.011\) & \(3.8 \%\)\\
\hline
vpu & \(248\) & \(\approx0.008\) &\( 2.7 \%\)\\
\hline
env & \(2570\) & \(\approx0.077\) & \(28 \%\)\\
\hline
nef & \(371\) & \(\approx0.012\) & \(4 \%\)\\
\hline
\end{tabular}\end{center}
\subsubsection*{Table 2:} The probability of spontaneous mutations in \textbf{RT}, \textbf{INT} and \textbf{PROT} genes of HIV-1, that are translated into the main target proteins of antiretroviral therapy.\\
\\
\begin{center}\begin{tabular}{|c|c|c|c|}
\hline
Gene & \((g)\) & \({P_g}\) & \% \({P_G}\)\\
\hline
RT & \(2078\) & \(\approx0.067\) & \(22.63 \%\)\\
\hline
INT & \(464\) & \(\approx0.015\) & \(5.05 \%\)\\
\hline
PROT & \(296\) & \(\approx0.01\) & \(3.23 \%\)\\
\hline
\end{tabular}\end{center}
\subsubsection*{Table 3:} Analysis of the probability of spontaneous mutations in structural genes \textbf{gag} and \textbf{env}.\\
\\
\begin{center}\begin{tabular}{|c|c|c|c|c|}
\hline
Gene & mPeptide & \((g)\) & \({P_g}\) & \% \({P_G}\)\\\hline
gag & P17 & \(452\) & \(\approx0.015\) & \(4.9 \%\)\\\cline{2-5}
    & P24 & \(692\) & \(\approx0.022\) & \(7.54 \%\)\\\cline{2-5}
    & P2 & \(41\) & \(\approx0.001\) & \(0.45 \%\)\\\cline{2-5}
    & P7 & \(164\) & \(\approx0.005\) & \(1.79 \%\)\\\cline{2-5}
    & P6 & \(155\) & \(\approx0.005\) & \(1.7 \%\)\\\hline
env & gp120 & \(1442\) & \(\approx0.047\) & \(15.7 \%\)\\\cline{2-5}
    & gp41 & \(1034\) & \(\approx0.034\) & \(11.26 \%\)\\\cline{2-5}
\hline
\end{tabular}\end{center}
\section*{DISCUSSION:}
Spontaneous mutation can arise from a variety of sources, and whatever the cause is, a large gene provides a large target and tend to mutate more frequently. Thus; the probability of spontaneous mutation is related to the ratio between the gene length and the whole genome length. This basic linear relation is used to formulate an equation that calculates the probability of emergence of spontaneous mutation in a certain gene per duplicate, depending on the ratio between the gene length and the whole genome length \(\frac{g}{G}\) in addition to the fixed genome-specific stability index \({S_i}\).\\

The drawbacks, which halt the development of HIV vaccines, are high mutability and variability of the virus. The mathematical analysis of each gene in HIV-1 genome (Table 1) indicates that \textbf{tat}, \textbf{vpr} and \textbf{vpu} are the least mutant genes per duplicate, so they are the best candidates for HIV-1 recombinant subunit vaccines or as a part of \textit{“prime and boost”} vaccine combinations. Also; the analysis indicates that the probability percent of spontaneous mutation emergence in the major three genes of HIV-1 genome \textbf{gag}, \textbf{pol} and \textbf{env} is 16.36 \%, 32.8 \% and 28 \% respectively. So; \textbf{pol} gene, which translated into polymerases enzymes, is the most susceptible gene for spontaneous mutations. Polymerases are currently the main targets for antiretroviral therapy and further analysis of \textbf{pol} gene indicates that reverse transcriptase \textbf{RT} gene is the most mutant among the polymerases.\\

The probability percent of spontaneous mutations in the \textbf{RT}, accounts for 22.63 \% of the total probability of spontaneous mutation emergence of the whole HIV-1 genome. Despite its high mutability; reverse transcriptase inhibitors should stay as a backbone of any highly active antiretroviral therapy (HAART). Reverse transcriptase, and due to its recombogenic properties and the absence of proofreading activity, is the core source of mutations in the HIV replication cycle. On the other hand; protease \textbf{PROT} gene is the least mutant in the polymerases. The probability percent of spontaneous mutations in the \textbf{PROT} accounts for 3.23 \% of the total probability of spontaneous mutation emergence of the whole HIV-1 genome (Table 2). Accordingly; protease inhibitors are better candidates, as a base, for antiretroviral combination therapy and the protease inhibitor-based regime represents a high genetic barrier for HIV to overcome.\\

The proposed mathematical analysis has many supportive clinical data. For example; The United Kingdom has one of the highest reported rates of primary resistance to HIV drugs worldwide. UK Group on Transmitted HIV Drug Resistance stated that the prevalence of resistance to any antiretroviral drug; to nucleoside or nucleotide reverse transcriptase inhibitors (NRTI), to non-nucleoside reverse transcriptase inhibitors (NNRTI), or to protease inhibitors (PI) were 19.2 \%, 12.4 \%, 8.1 \%, and 6.6 \%, respectively [5]. In Spain; a study stated that the prevalence was 5.8 \% for NRTI, 5 \% for NNRTI and 3.8 \% for PI [6]. In Turkey; a study stated that the percentage of HIV-1 primary drug resistance mutations, in antiretroviral therapy-naive patients, was 5.2 \% for NRTI, 3.1 \% for NNRTI and 2.1 \% for PI [7]. In Djibouti; a study indicated that among 16 patients with first-line ART failure, 56.2 \% showed reverse transcriptase inhibitor-resistant HIV-1 strains. But on the contrary; no protease inhibitor resistant strains were detected [8]. All these findings indicate that resistance emergence to protease inhibitors is much lesser than that of reverse transcriptase inhibitors.\\

In a wider scope; the main advantage of the proposed mathematical approach is providing a linear equation to calculate the probability of spontaneous mutation per duplicate for simpler viral genomes. Otherwise; further analysis is needed before recruiting this equation to make a mutability map for more complicated bacterial or eukaryotic genomes. If the equation is applicable on these more complex genomes, it will indicate that noncoding genome segments, which present in the genome of prokaryotes and eukaryotes by different portions, not only perform regulatory functions, but also protect the genetic information of the coding genome by providing a wider genetic pool.\\

Moreover; the proposed equation is useful for antiviral drug activity interpretation, as the mutability of the targeted protein plays an integral role in determining in vivo drug activity. Furthermore; the equation provides a general picture about the mutability of each gene in a targeted viral genome.  This can be helpful during drug development researches, and during protocoling of combination therapy. The developers can target proteins translated from the relatively lesser mutating genes.\\

On the other hand; the main disadvantage of the proposed equation is numerical bias during expressing it with numerical values. As an example; the numerical value of the stability index for HIV-1 genome can be biased. \textit{Mansky and Temin} reported that the forward mutation rate for HIV-1 was \(3.4\times10^{-5}\) mutations per bp per cycle [9], while \textit{Cuevas et al} used the intrapatient frequency of premature stop codons to quantify the HIV-1 genome-wide rate of spontaneous mutation in DNA sequences from peripheral blood mononuclear cells, which revealed a mutation rate of \((4.1\pm1.7)\times10^{-3}\) per base per cell [10].\\

In addition to the fact that the emergence of antiviral drug resistance is a multifactorial process [11]; the proposed equation only provides the probability of spontaneous mutation emergence in a specific gene, but does not determine which of these emergent mutations are lethal and which are not. The lethally mutated viral genomes fail to reach the plasma leading to mass deletion of the emergent mutation.\\
\section*{CONCLUSION:}
The mathematical analysis of HIV-1 genome indicates that \textbf{tat}, \textbf{vpr} and \textbf{vpu} are the least mutant genes per duplicate, so they are the best candidates for HIV-1 recombinant subunit vaccines or as a part of \textit{“prime and boost”} vaccine combinations. Also; protease inhibitors are better candidates, as a base, for antiretroviral combination therapy and the protease inhibitor-based regime represents a high genetic barrier for HIV to overcome. In a wider scope; The proposed equation offers a wider array of options for drug developers and during drug combination protocoling to help predict in vivo antiviral drug activity and to deal with mutation-induced drug resistance.\\
\section*{ACKNOLEDGEMENT:}
The author states that there is no conflict of interest regarding this article.\\
\section*{REFERENCES:}
\begin{enumerate}
\item Rambaut, Andrew, et al. "The causes and consequences of HIV evolution." Nature Reviews Genetics 5.1 (2004): 52-61.\\
\item Davies, Julian, and Dorothy Davies. "Origins and evolution of antibiotic resistance." Microbiology and molecular biology reviews 74.3 (2010): 417-433.\\
\item Robertson, David L., Beatrice H. Hahn, and Paul M. Sharp. "Recombination in AIDS viruses." Journal of molecular evolution 40.3 (1995): 249-259.\\
\item Petropoulos, C. "Retroviral taxonomy, protein structures, sequences, and genetic maps." Retroviruses. Cold Spring Harbor Laboratory Press, Cold Spring Harbor, NY (1997): 757-805.\\
\item Cane, P., et al. "Time trends in primary resistance to HIV drugs in the United Kingdom: multicentre observational study." BMJ (Clinical research ed.) 331.7529 (2005): 1368-1368.\\
\item Yebra, Gonzalo, and Africa Holguin. "Epidemiology of drug-resistant HIV-1 transmission in naive patients in Spain." Medicina clinica 135.12 (2010): 561-567.\\
\item Yalçınkaya, T., and S. Köse. "Investigation of HIV-1 primary drug resistance mutations in antiretroviral therapy-naive cases." Mikrobiyoloji bulteni 48.4 (2014): 585-595.\\
\item Abar, Aden Elmi, et al. "HIV-1 drug resistance genotyping from antiretroviral therapy (ART) naïve and first-line treatment failures in Djiboutian patients." Diagnostic pathology 7.1 (2012): 138.\\
\item Mansky, Louis M., and Howard M. Temin. "Lower in vivo mutation rate of human immunodeficiency virus type 1 than that predicted from the fidelity of purified reverse transcriptase." Journal of virology 69.8 (1995): 5087-5094.\\
\item Cuevas, José M., et al. "Extremely high mutation rate of HIV-1 in vivo." PLoS Biol 13.9 (2015): e1002251.\\
\item Paydary, Koosha, et al. "The emergence of drug resistant HIV variants and novel anti–retroviral therapy." Asian Pacific journal of tropical biomedicine 3.7 (2013): 515-522.\\
\end{enumerate}
\end{document}